# High Performance Simulation of Spaceborne Radar for Remote-Sensing Oceanography: Application to an Altimetry Scenario

Goulven Monnier
Scalian DS
2 rue Antoine Becquerel,
35700 Rennes, France
goulven.monnier@scalian.com

Benjamin Camus
Scalian DS
2 rue Antoine Becquerel
35700 Rennes, France
benjamin.camus@scalian.com

Yann-Hervé Hellouvry
Scalian DS
2 rue Antoine Becquerel
35700 Rennes, France
yann-herve.hellouvry@scalian.com

*Abstract*—In this paper, we detail the high-performance implementation of our spaceborne radar simulator for satellite oceanography. Our software simulates the sea surface and the signal to imitate, as far as possible, the measurement process, starting from its lowest level mechanisms. In this perspective, raw data are computed as the sum of many illuminated scatterers, whose time-evolving properties are related to the surface roughness, topography, and kinematics. To achieve efficient performance, we intensively use GPU computing. Moreover, we propose a fast simulation mode based on the assumption that the instantaneous Doppler spectrum within a range gate varies on a timescale significantly larger than the PRI. The sea surface can then be updated at a frequency much smaller than the PRF, drastically reducing the computational cost. When the surface is updated, Doppler spectra are computed for all range gates. Signals segments are then obtained through 1D inverse Fourier transforms and pondered to ensure a smooth time evolution between surface updates. We validate this fast simulation mode with a radar altimeter simulation case of the Sentinel-3 SRAL instrument, showing that simulated raw data can be focused and retrieved using state-of-the-art algorithms. Finally, we show that, using a modest hardware configuration, our simulator can generate enough data in one day to compute the SWH and SSH spectra of a scene. This demonstrate that we achieved an important state-of-the-art speed-up.

## I. Introduction

Simulations are extensively used in the field of earth observation from space, for as various purposes as mission performance assessment and planification, processor prototyping, help to the understanding of observations and to testing hypotheses on involved mechanisms or artificial intelligence training. Usually, no simulator has the ability to reproduce all levels of processes and events involved in a mission, so that several specialized and complementary simulators may be required. Among them, end-to-end (E2E) simulators are in charge of the lowest levels. They include a forward model comprising environment, acquisition geometry and instrument modules (designed to produce synthetic realistic raw data) and ground processors prototypes.

SCALIAN has been developing for years a raw data generator for satellite oceanographic radars, which is now an essential component of the ESA Harmony mission E2E simulator. This paper describes an application to radar altimetry carried out by SCALIAN and CLS under CNES R&T funding. The goal of the study is to investigate, thanks to simulated data, the origin of the "bump" widely observed in altimetric sea surface altitude (SSA) and significant wave height (SWH) spectra, in low resolution (LRM) and SAR modes [1]. The simulated instrument is SRAL, the nadir SAR radar altimeter on board Sentinel-3 satellites operating in Ku band (13.575 GHz).

Whatever the considered radar instrument, generating synthetic raw data is very expensive because of large illuminated ocean areas having to be simulated at metric resolution and high time rate. In the case of the present study, this is even made worse because of SRAL high PRF of 18 kHz (which must actually be simulated for SAR processing) and, above all, because of the need for retrieving time spectra along the satellite track. Obtaining one sufficiently clean and resolved spectrum typically requires averaging over ~100 realizations of acquisition along a ~100 km satellite track. Despite several algorithmic optimizations and a multi-GPU implementation, producing ~20-30 spectra as required for the study was not affordable. We thus designed and implemented a new "fast mode" allowing reducing considerably the time rate of the heaviest parts of the simulation while preserving the key properties of the generated IQ signals.

In the following, the simulator principles and architecture are briefly presented, then the paper focuses on the fast mode algorithm and implementation. Sample results are finally presented and discussed.

## II. Simulator

### A. Overview

In a non-classical approach, we merge the scene and the instrument parts to imitate, as far as possible, the measurement process, starting from its lowest level mechanisms. In this perspective, raw data are computed as the sum of a huge number of illuminated scatters, whose time-evolving scattering properties are related to the surface topography and kinematics, which is the only practical approach which allows to account for all influential characteristics of the instrument at the same time. The whole approach is illustrated in Figure 1.

Hence, the module will simulate the sea surface to fit the images to be simulated, in terms of area and resolution. In a regular mode, this surface will be updated at the PRF to precisely reproduce the surface kinematics and its Doppler contribution. The surface simulation strategy will consist of in generating surrogate 3D surfaces from prescribed spectra, which is the only practical method. We rely on the Choppy Wave Model [2] to introduce wave profiles asymmetry and resulting non-gaussian statistics.

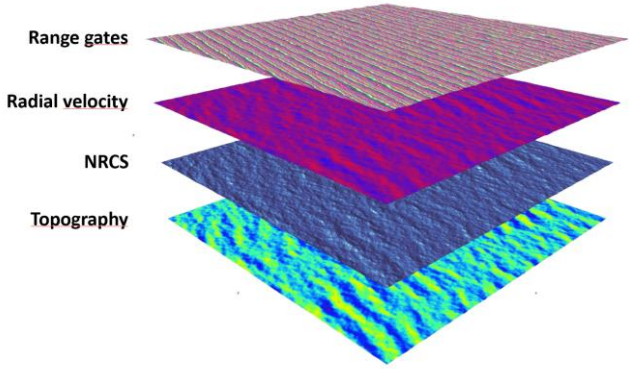

*Figure 1. Visualization of the different simulation steps. (1) Generation of the sea surface. (2) Computing the NRCS of the scatterers (3) Doppler simulation through surface animation (4) complex summation of the scatterers in range gate.*

To model the scattering of an incoming EM wave by such an extended, gridded, moving sea surface, we rely on the "Two Scale Grid Model" (TSGM), which consists in separating resolved and unresolved scales under radar wavelength, surface roughness, and numerical sampling considerations. Offline, we make use of the SSA1 [3] model to compute look-up tables of the NRCS and the Doppler of unresolved waves as a function of sea state and local scattering geometry parameters. The received instantaneous backscattered signal of the resolved waves is essentially obtained by summing the complex contributions from all the facets lying in each range gate.

During the generation of raw data, all involved antenna patterns are estimated for each pulse at each contributing facet. Thanks to a pre-processing step, all antenna patterns required for a simulation are computed and recorded on a u-v grid. The complex values of all antenna patterns at each facet are then obtained through interpolation in the u-v grid. The radar equation is applied at the facet level in terms of complex amplitudes. More details on the implementation are given in the next section.

*B. Implementation Detail*

The simulation requires a large number of resources and is computationally consuming. To optimize and make the chain scalable with very large scenes, a parallelisation approach is adopted relying on a general-purpose processing on graphics processing units (GPGPU). Due to the limited memory available on these accelerators, we partition the scene, i.e. the facets information (vertex, faces, normals), into multiple batches, called tiles here after. The tiles are computed sequentially on the GPU. Then, for each tile and radar pulses, the corresponding surface is generated and the radar equation is applied to compute the scatterers impulse response.

If available on the computing node, the computation can also be distributed on several devices. In this multi-GPU strategy, the available devices divide up the batch of tiles to process. The different tiles are then simulated in parallel on different accelerators. Therefore, each GPU computes only a subpart of the final raw data. Once all the chunks have been processed, a final thread gathers the sparse raw data. This implementation is illustrated in Figure 2.

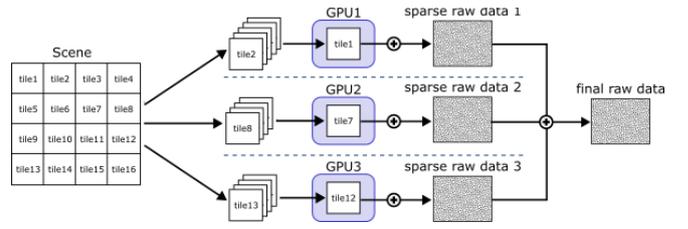

*Figure 2. Illustration of the multi-GPU implementation.*

### III. FAST SIMULATION MODE

The fast simulation mode we propose assumes the instantaneous Doppler spectrum of each range gate to vary at a timescale significantly larger than the PRI. This assumption should hold for most radar system because these conditions are required anyway to ensure Doppler measurements and consistency. Based on this assumption, the sea surface is updated at a frequency significantly smaller than the PRF, which drastically reduces the computation cost. In the following, we first detail the algorithm, and then its GPU implementation.

*A. Algorithm*

With the fast simulation mode, we do not directly compute the radar signal from the updated sea surface for all the pulses. Instead, we compute every $n_{fast}$ pulses, the instantaneous Doppler spectrum of each range gate. To this end, for two successive pulses $p_i$ and $p_{i+1}$, we update the sea surface, and evaluate the complex amplitude of the scatterers. Then, we determine the Doppler frequency of the scatterers thanks to the pulse-pair method. Finally, we sum the scatterers according to their range gate and Doppler interval. All the spectra are then converted through inverse 1D Fourier transforms to long time signals centered on the current pulse (i.e. signals that propagate to the past and future pulses). To ensure continuity of the data, the signal obtained from a given Doppler spectrum is combined through weighted sum with the signals obtained from the previous and the following Doppler spectra (in the same range gate). This strategy is illustrated in Figure 3.

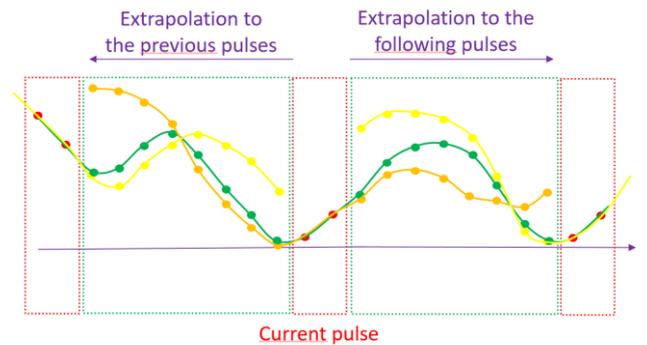

*Figure 3. Illustration of the long-time signal combination. The red dots correspond to the pair of pulses used to compute the Doppler spectra. The signal obtain from the current Doppler spectrum is in orange. The signal from the past and future Doppler spectra are in yellow. The signal resulting from the combination of the past, current and future signals is in green.*

The weights follow the square root of a linear function of the pulse coordinate:

$$W_i(x) = \left(1 - \frac{|i-x|}{n_{fast}}\right)^{1/2} \quad (1)$$

With:

- $i$ the index of the pulse corresponding to the Doppler spectrum
- $x$ the index of the pulse

The ½ power in $W_i$ ensures that the total variance is conserved, on average.

*B. Implementation*

We implemented this fast-mode scheme in the following way. First, a 3D tensor $T_{fast}$ of dimensions $(1 + \frac{n_{pulses}}{n_{fast}}, n_{range}, n_{doppler})$ is allocated in memory to contain all the Doppler spectra required for the simulation. Then, all the Doppler spectra are computed on GPU according to Algorithm 1 (with $n_{pulses}$ the total number of pulses in the simulation).

```
For i in [0, n_pulses/n_fast] do
    update_simulation(i * n_fast/PRF)
    C_i ← scatterers_complex_amplitudes()
    R_i ← scatterers_range()
    update_simulation((i*n_fast+1)/PRF)
    C_{i+1} ← scatterers_complex_amplitudes()
    D_i ← arg(C*_{i+1} C_i) PRF
    rasterize(C_i, T_fast, i, R_i, D_i)
```

*Algorithm 1. Computation of all the Doppler spectra during the fast-mode simulation.*

This algorithm uses the following methods:

- `update_simulation(t)`: updates the sea surface, and the instrument position at time $t$ (in seconds)
- `scatterers_complex_amplitudes()`: computes and return the complex amplitudes of the scatterers for the current simulation state
- `rasterize(D,T,x,y,z)`: sums the data $D$ in the tensor $T$ according to the $x$, $y$, and $z$ coordinates

Once this phase is finished, a post-processing step is in charge of converting the Doppler spectra to the final range-azimuth IQ results. This step consists of:

1. Allocating the memory for the final range-azimuth IQ result matrix $Y$
2. Running batch of 1D iFFT on the last dimension (i.e. the Doppler dimension) of $T_{fast}$.
3. Weighting the obtain time-long signals in the last dimension using equation (1).
4. Summing the weighted time-long signals in the corresponding cells of $Y$.

Sophisticated CPU and GPU memory management have been developed to accommodate the GPU memory limits while ensuring that large IQ data can be simulated. The main data structures are allocated in CPU RAM while the GPU memory only contains smaller caches. When computing the Doppler spectrum, a GPU cache tensor is progressively filled. As soon as the cache is full, the data are copied to the CPU RAM and summed in the correct cells of the complete tensor. The cache is finally cleared to continue the processing.

During the post-processing phase, only a part of the CPU tensor is copied on the GPU cache. The post-processing step (i.e. iFFT and weighted sum) is then performed in the GPU to store the final range-azimuth results in a cache matrix. Finally, this cache is copied in CPU in the correct cells of the complete matrix. The following part of the tensor is then copied on the GPU, and the whole process is repeated until we processed all the results.

## IV. VALIDATION WITH AN ALTIMETRY SCENARIO

To validate our approach, we simulate an altimetry acquisition with the SRAL instrument of the Sentinel-3 mission. We consider a trajectory that span a 100 km long area with a PRF of 17.8 kHz. The simulation comprises a total of 267 379 pulses. In the simulation we consider scatterers of 6.24m² and simulation tiles of 1024x1024 scatterers. The simulation requires 335 tiles for a total of 3.51e8 scatterers the dynamically move at each radar pulses.

We ran all the simulations on a computing node with 4 GPUs Nvidia GeForce GTX 1080, which is a modest hardware configuration compared to traditional high-performance computing infrastructures. We tested several values of $n_{fast}$ between 0 (i.e. the regular simulation mode) and 1250 by increment of 250. In the following, we first present the execution performance, then we analyze the data in LRM and Delay-Doppler modes.

*A. Performance measurements*

To run a representative use-case we simulated a sea with a wind of 9 m/s and a swell with an SWH of 4m. We also enable the choppy wave model. Despite a massive parallelization of the computation, the simulation took 20 hours and 10 minutes to execute in the regular mode. With a single GPU it took 60h. This duration may seem very long, but it must be put in perspective with the important size of the simulated scene. As shown in Figure *4*, with our fast mode, the simulation took only 20 minutes with $n_{fast} = 250$, which represents an important x60 speed-up. We observe that this duration decreases significantly when $n_{fast}$ increases. With $n_{fast} = 500$, the simulation is 87 times faster than with the regular mode as it ends after only 14 minutes. For $n_{fast} \geq 1000$, the execution time reach an asymptote at 11 minutes for a speed-up factor of 114. This asymptote is caused by static simulation cost such as CPU/GPU memory allocation, disk access. To sums up, we show here that our fast mode achieves an important speed-up.

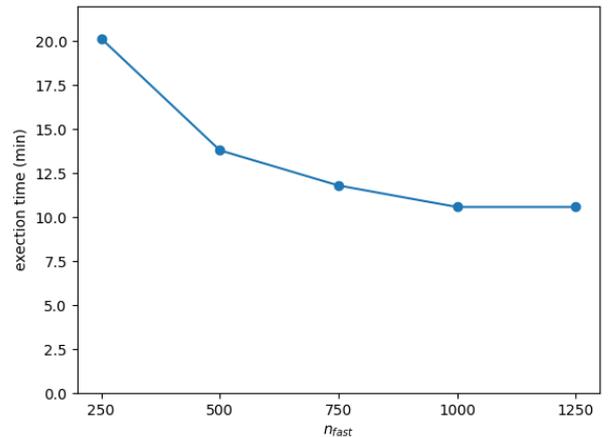

*Figure 4. Execution time of the simulation for our fast mode.*

## B. LRM data analysis

To validate the simulated data, we first analyze the LRM profiles. We disabled the choppy wave model here to remove the bias induced by a skewed heigh distribution in the altimetry echo. We applied a multi-look of the pulses over a 50ms period to reduce the speckle. The Figure 5 shows the results obtained with $n_{fast} = 500$.

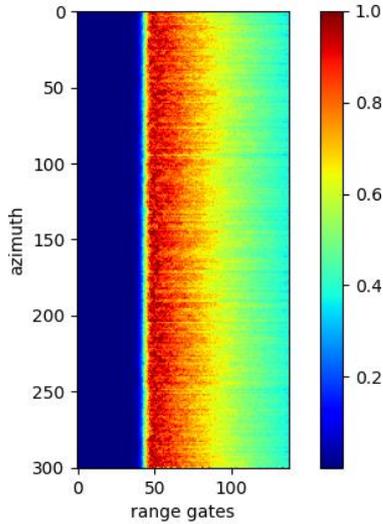

*Figure 5. Visualization of the multi-looked normalized power simulated LRM data.*

We observe from Figure 6 that the mean normalized power waveforms obtained with the regular and fast modes are similar for all the values of $n_{fast}$ tested. This indicates that the approximations introduced by the fast mode remains valid even for large value of $n_{fast}$.

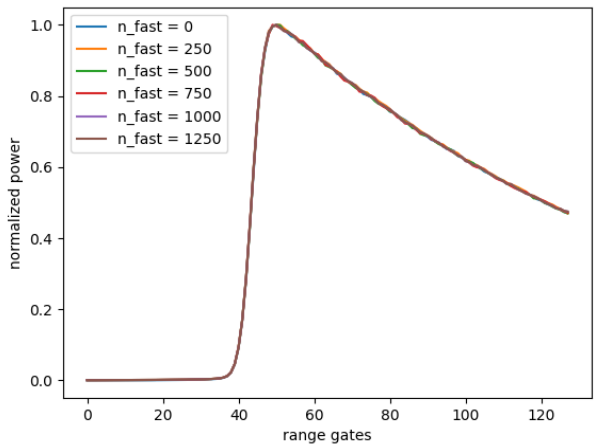

*Figure 6. Comparison of the mean power waveforms of the LRM simulated data obtained with different value of $n_{fast}$.*

To show that the simulation waveforms are valid, we go one step further and apply a retracking algorithm to retrieve the SWH of the scene from the LRM signal. We used the MLE4 algorithm that consist of fitting a second-order analytic model of the waveform on each echo thanks to a maximum likelihood estimation [4]. The ocean parameters are then directly derived from the learned parameters of the model. For each value of $n_{fast}$, we simulated different scenes with U10 speed of 5, 10, 15, and 20 m/s, without a swell. We compare then the ground truth SWH to the predicted one.

The results are shown in Figure 7. We globally observe that the predictions are in good accordance with the ground truth. The highest prediction errors (of about 28 cm) are met when the SWH is smaller than 1m, which correspond to an unusual, noisy scenario. For higher SWH, we observe a mean prediction error of 8 cm that is typical of the retracking algorithms accuracy [5, 6]. Correction LUT can be defined to compensate for this error, (which is mainly caused by the gaussian approximation of the Point Target Response (PTR) made during the retracking [7]) but this is out of the scope of this article. Nevertheless, we found that the retrieved SWH are similar with the fast mode and the regular one (i.e. $n_{fast} = 0$) with only 2 cm of difference in average. As, these small variations are significantly smaller than the retracking algorithm accuracy, this demonstrate that the approximations introduced by the fast mode are valid in the altimetry context. In LRM, where, contrary to SAR modes, the signal phase is not used, this essentially shows that the speckle noise affects the retracking the same way in normal and fast mode, up to high values of $n_{fast}$. Hence, the time correlation of the IQ signals remains correctly reproduced in fast mode.

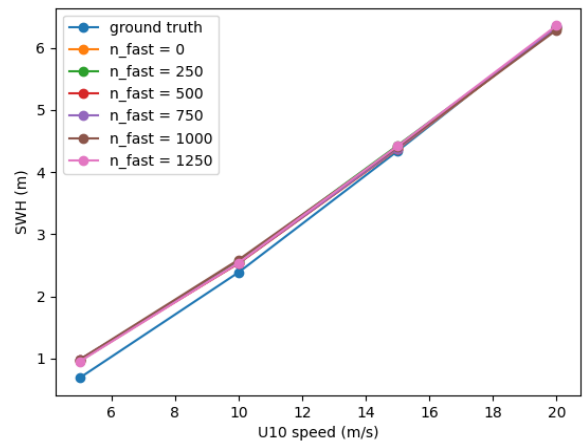

*Figure 7. Comparison of the ground truth SWH with the one retrieved from the LRM data for different value of $n_{fast}$ and U10 speed.*

## C. Delay-Doppler data analysis

To show that the simulated data generated with our fast-mode can be focused, we applied a Delay-Doppler algorithm [8]. It consists of splitting the data into bursts of 64 pulses, and for each burst performing:

1. a correction of the doppler centroid shift to compensate for the instrument trajectory that is not perfectly tangent to the surface [9].
2. an along-track FFT to increase the azimuth resolution up to 300m.
3. a correction of the scatterers range migration with the Doppler through the application of delay phase coefficients.

Multi-look is then applied by averaging the Doppler strips to reduce the speckle.

The Figure 9 shows that the results obtained with $n_{fast} = 500$ is well focused. Moreover, as shown on Figure 9, we observe that the normalized mean power range profile matches the typical Delay-Doppler waveform. We see that this profile is stable for all the $n_{fast}$ values tested and looks like the one obtained with the regular mode.

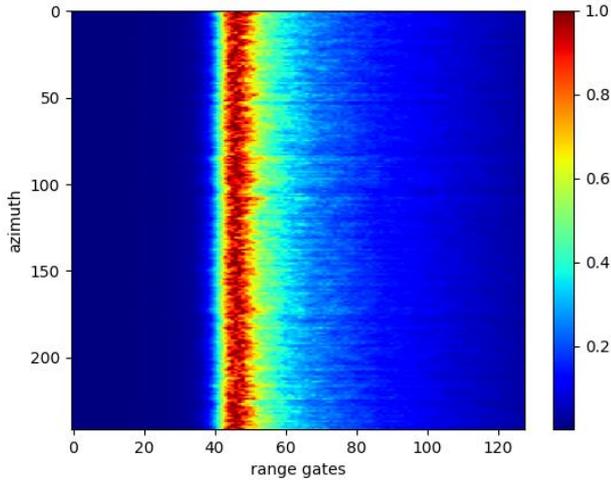

Figure 8. Simulated raw data processed with the Delay-Doppler algorithm.

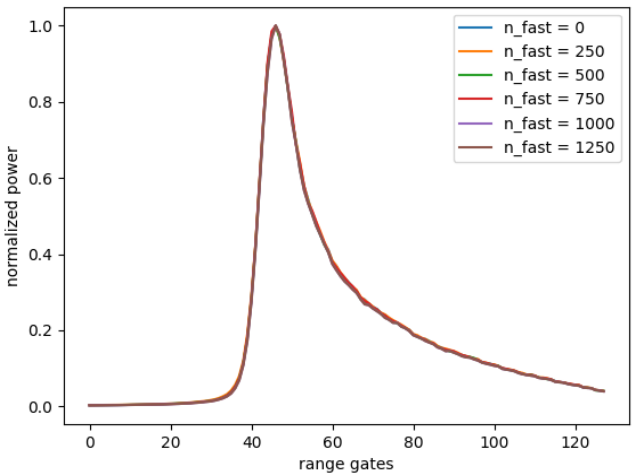

Figure 9. Comparison of the mean power waveforms of the simulated data obtained with different value of $n_{fast}$ and processed with the Delay-Doppler algorithm.

Like with the LRM data, we go one step further and retrieve the SWH of the scene from the Delay-Doppler data using the DDA4 retracking algorithm [10]. It consists of fitting a semi-analytic model of the Delay-Doppler waveform on each focused echo using the Levenberg-Marquardt algorithm. As shown in Figure 10, we observe that the retrieved SWH are close to the ground truth with absolute gaps ranging from 1 cm to 16 cm, for an average of 6 cm. Importantly, for all the value of $n_{fast}$ tested, the accuracy of the predictions remains similar (event sometimes better) than with the regular simulation mode. This demonstrates that the approximation of the fast mode remains valid for the Delay-Doppler process even with large $n_{fast}$ values.

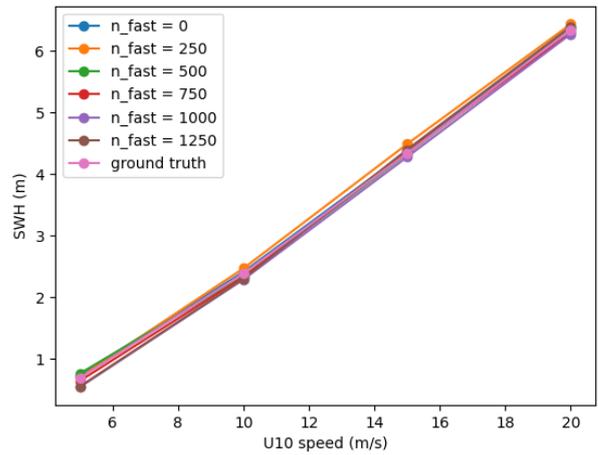

Figure 10. Comparison of the ground truth SWH with the one retrieved from the LRM data for different value of $n_{fast}$ and U10 speed.

D. SWH and SSH spectra computing

The high computing performance of our fast mode unlock new experimental capabilities. To demonstrate this point, we ran enough simulation to compute the SWH and SSH spectra of a scene. Setting $n_{fast} = 500$, we performed 100 simulation runs (with different sea surfaces instantiation for each run) in less than 1 day.

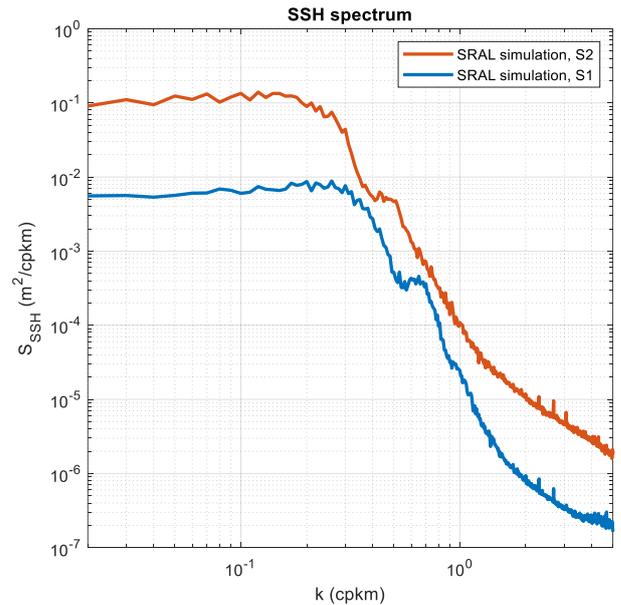

Figure 11: SSH spectra obtained for sea states S1 and S2

Figure 11 shows two spectra, each obtained from 100 realizations of the altimetric LRM measurement along a 100 km track, for the following gaussian swell spectra:

- S1: $SWH_{swell} = 2.5\ m$, $\sigma_{k_x} = \sigma_{k_y} = 0.006\ m^{-1}$
- S2: $SWH_{swell} = 5\ m$, $\sigma_{k_x} = \sigma_{k_y} = 0.003\ m^{-1}$

Where $SWH_{swell}$ is the swell SWH and $\sigma_{k_x}$, $\sigma_{k_y}$ the gaussian stds in along- and cross-track directions. In both cases the wind is oriented along-track with $U_{10} = 7\ m/s$. In the absence of other sources of SSH variations in the simulation, the observed spectra can be attributed to the swell and prove to reproduce the main characteristics of the spectral bumps seen in real measurements. The frequency cutoff is related to the footprint of the instrument in LRM mode, which is slightly dependent on the SWH.

Figure 12 illustrates how LRM (MLE4) and SAR (DDA4) spectra compare, for SWH and SSH. The SWH spectra get close to each other at low frequency, as both LRM and SAR footprints become small compared to wave groups length. On the high-frequency side, the SAR cutoff takes place at higher frequency, because of its much smaller footprint. The SAR SSH spectrum is about a decade lower than the LRM spectrum, in accordance with observational data. This is also a result of reduced footprint in the along-track direction. Beyond this qualitative interpretation, such numerical simulations are currently used in support to a modelling effort based on expressing the modulation transfer function relating the estimated SSH (and SWH) to spatially varying SWH due to wave groups.

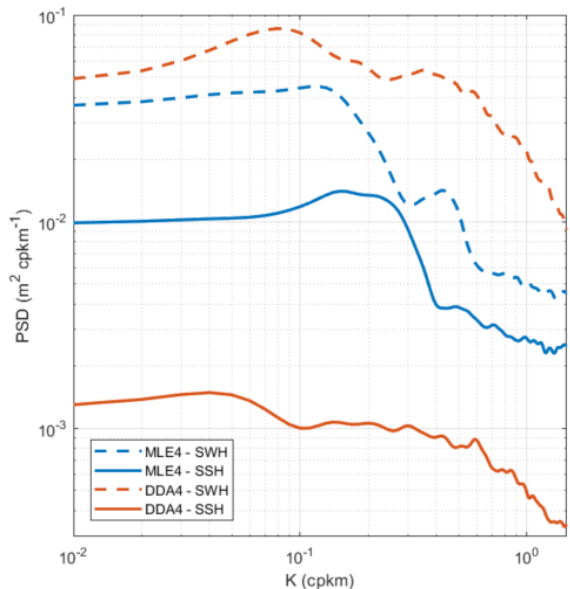

*Figure 12. SWH and SSH spectra for LRM and Delay-Doppler data.*

## V. Conclusion

In this paper, we detailed the high-performance implementation of our spaceborne radar simulator for remote-sensing oceanography. This implementation heavily relies on multiple GPU accelerators to massively parallelize the simulation. We proposed here a fast mode computing scheme that consist of updating the sea surface at a frequency much smaller than the PRF, drastically reducing the computational cost. When the surface is updated, Doppler spectra are computed for all range gates. Signals segments are then obtained through 1D inverse Fourier transforms and pondered to ensure a smooth time evolution between surface updates. Using an altimetry scenario with the SRAL instrument of Sentinel-3, we demonstrated the efficiency, validity and robustness of this fast mode. We show that the fast mode results are similar than the ones obtained with regular simulations, and that the raw data can be processed using state-of-the-art algorithms such as focusing, and retracking. With a speed-up factor of more than 100, the simulations of large scene of more than 3.5e8 dynamically moving scatters illuminated by 267 379 pulses ends in about 10 minutes on a modest hardware configure of 4 regular GPUs. We have shown that this enables to conduct computationally demanding experiments such as determining SWH and SSH spectra.

Our experiments shows that we can use large value of the $n_{fast}$ (i.e. the number of skipped pulses between two surface update) in the case of the altimetry to maximize the speed-up. This may not be the case in another context such as off-nadir SAR imaging where the PRF are significantly smaller. More experiments are needed to calibrate our fast mode in such scenario. We expect $n_{fast}$ to scale with the PRF. This is interesting because, other things being equals, scenarios with high PRF are precisely the ones that have the highest computation costs, as they require to simulate more pulses. This means that our fast mode should deliver higher speed-up on the slowest simulation. Once again, more experiments are required to quantify this phenomenon.

Thanks to our fast mode, we are in capacity generating sufficient labelled data to train inversion models thanks to Deep Learning algorithms. This may unlock new possibilities for remote sensing-oceanography.

ACKNOWLEDGEMENTS

This work has been partially funded by the CNES in the context of the "Caractérisation des erreurs méso-échelle du mode altimétrie Doppler et stratégie de correction" R-S21/OT-0003-093 R&T.